\PassOptionsToPackage{hypertexnames=false}{hyperref}
\documentclass[submission]{eptcs}
\providecommand{\event}{FMAS 2026} 

\usepackage{underscore}         

\usepackage[T1]{fontenc}
\usepackage[utf8]{inputenc}
\usepackage{graphicx}
\usepackage{amsmath,amssymb,amsfonts}
\usepackage{hyperref}
\usepackage{cleveref}
\usepackage{booktabs}
\usepackage{float}       
\usepackage{microtype}  
\usepackage{color}
\usepackage[disable]{todonotes}
\usepackage{lineno}
\usepackage{tikz}
\usepackage{xurl}
\usepackage{listings}
\lstdefinelanguage{SLIM}{
	keywords={system,implementation,properties,end, Globally, eventually, continually},
	keywordstyle=\bfseries\color{blue!65!black},
	morekeywords=[2]{Patterns,Name,Pattern},
	keywordstyle=[2]{\bfseries\color{teal!80!black}},
	morekeywords=[3]{if, has, occurred, then, in, response, holds},
	keywordstyle=[3]{\itshape\color{purple!90!black}},
	sensitive=true,
	morecomment=[l]{--},
	commentstyle=\itshape\color{gray!70},
}
\usetikzlibrary{arrows.meta,automata, positioning, fit, positioning, backgrounds, calc, shapes.misc, shapes.geometric}

\hypersetup{
	hidelinks,
	colorlinks=true, 
	citecolor=blue, 
	linkcolor=blue, 
	urlcolor=blue,  
	pdfauthor={},
	pdftitle={},
}

\title{Formal Analysis of a Fault-Tolerant Extension of the ROSACE
  Flight Controller}
		
\author{Marco Bozzano
  \institute{Fondazione Bruno Kessler, Trento, Italy}
  \email{bozzano@fbk.eu} \and
Alessandro Cimatti
  \institute{Fondazione Bruno Kessler, Trento, Italy}
  \email{cimatti@fbk.eu} \and
Moussa Fall
\institute{Fondazione Bruno Kessler, Trento, Italy}
\institute{University of Trento, Trento, Italy}
  \email{mfall@fbk.eu} \and
Stefano Tonetta
  \institute{Fondazione Bruno Kessler, Trento, Italy}
  \email{tonettas@fbk.eu}
}
\def\titlerunning{Formal Analysis of the ROSACE Flight Controller}
\def\authorrunning{M. Bozzano et al.}
\begin{document}
	
\maketitle
	
\begin{abstract}
Autonomous avionics controllers must satisfy stringent real-time
functional and performance requirements even in the presence of
run-time faults. In this paper, we analyze ROSACE, an open-source
control engineering benchmark for flight control systems, focusing on
its run-time resilience against faults such as sensor noise and sensor
failures. We extend the benchmark with an FDIR (Fault Detection,
Identification and Recovery) component based on Triple Modular
Redundancy.

		
Our approach builds upon COMPASTA, a tool for Model-Based Design
developed on top of the TASTE tool by ESA. COMPASTA enables the
modeling and injection of faults into a system's design. Moreover, it
implements a simulation-based verification and validation workflow,
automating the generation and execution of a set of test cases, based
on a user-defined notion of coverage.  Using COMPASTA, we construct a
set of formal models, with faults and with FDIR, and we formally
analyze their impact on the system properties. The analysis is based
on the coupling of the formal models with the Simulink models used for
the simulations.
		
\end{abstract}
	
\section{Introduction}
\label{sec:intro}
The ROSACE case study~\cite{pagetti2014rosace} is an open-source
avionics benchmark focused on the design and implementation of a
longitudinal flight controller for a civil aircraft. It models the
autonomous operation of an aircraft during cruise and step-climb
maneuvers, where the autopilot regulates altitude, vertical speed, and
airspeed. The case study is particularly complex because it combines
continuous aircraft dynamics with discrete-time digital control laws
and heterogeneous execution rates. Another source of complexity comes
from balancing control performance requirements (stability, rise time,
overshoot, settling time) with real-time execution
constraints. In~\cite{pagetti2014rosace}, the system was designed and
analyzed in Matlab/Simulink.

In this work, we investigate autonomy properties of the ROSACE
avionics controller, focusing on its ability to autonomously react to
anomalies and disturbances, while preserving formally specified safety
properties. Specifically, we extend ROSACE by introducing run-time
faults, such as sensor noise and sensor failure, and we analyze their
impact on the system requirements. Our first contribution is the
formalization of the requirements, which are expressed in natural
language in~\cite{pagetti2014rosace}, in temporal logic. Second, we
use an approach based on formal safety assessment~\cite{xsap-journal}
to model faults and their effects, and to automatically inject them
into the system model. Third, we systematically verify the safety and
performance requirements under different scenarios, using a
simulation-based formal verification workflow, called
VIVAS~\cite{ISoLA24}. Namely, scenarios of interest are automatically
generated from a formal model of the system, concretized and
simulated. Scenario generation is carried out using model checking
techniques, which guarantee that the generated test cases meet
appropriate coverage criteria. Formal models are co-simulated along
with (variants of) the original Simulink model, and system
requirements are formally verified on the resulting execution traces.
Finally, we further extend ROSACE by modeling sensor redundancy and a
simple FDIR~\cite{savoir} (Fault Detection, Identification and
Recovery) sub-system based on a TMR (Triple Modular Redundancy)
architecture~\cite{flad2026comprehensivesurveyredundancysystems} with
a simple voting mechanism to reach a fault-tolerant consensus. We
compare the original model against the model variants with faults,
with and without FDIR, and we analyze the impact on the safety and
performance of the system.

Our approach builds upon
COMPASTA~\cite{IMBSA22,AUJ23,CEAS24}\footnote{COMPASTA is available at
\url{https://gitlab.esa.int/taste/taste-setup/-/tree/fbk-update-compasta-toolchain/misc/compasta-code?ref_type=heads}. Merge
with the main development branch of TASTE is planned as future
work. An archive containing all the artifacts is available at the
following link:
\href{https://es-static-lfs.fbk.eu/access/bozzano/people-bozzano/extended-rosace-experiments-main.zip}{extended-rosace-experiments-main.zip}.}
, a tool for model-based design developed on top of the TASTE tool by
ESA~\cite{TASTE2}. TASTE is a Model Based Systems Engineering (MBSE)
tool and development environment for embedded, real-time systems,
which has been actively developed by ESA since 2008, offering
capabilities such as automatic code generation, deployment and
simulation. With respect to TASTE, COMPASTA enables the modeling of SW
along with HW components and their faults, using languages such as
SDL~\cite{SDL} and AADL~\cite{AADL}. It supports automatic fault
injection, formal verification and safety assessment, based on model
checking techniques. Moreover, it implements the VIVAS workflow, and
it supports the modeling of AI-based components~\cite{ISoLA24}.

We report our experience in modeling and verification of the extended
version of the ROSACE controller. We discuss the results of our
analyses in detail, compare them with the original results in
~\cite{pagetti2014rosace}, and evaluate the impact of faults and FDIR
on the satisfaction of system requirements. We discuss advantages of
using formal methods and lessons learnt. In particular, we show that
using COMPASTA and the VIVAS workflow offers clear advantages over
plain modeling in Matlab/Simulink, in terms of the systematicity of
the analyses performed and the quality of the results. Notably,
automated generation of test scenarios based on a formal model,
enables tailoring of the generated test cases with respect to a
user-defined notion of coverage, and a-posteriori verification of the
simulated scenarios provides a formal guarantee on the actual coverage
of the analyzed scenarios.

The rest of this paper is organized as follows. In
Sect.~\ref{sec:related}, we discuss related work. In
Sect.~\ref{sec:rosace}, we describe the ROSACE case study. In
Sect.~\ref{sec:modeling-simulink}, we present the modeling of the case
study and its extensions in Matlab/Simulink, while in
Sect.~\ref{sec:modeling-compasta}, we present the modeling in
COMPASTA. In Sect.~\ref{sec:analysis}, we discuss the analysis of the
case study.  In Sect.~\ref{sec:experimental} we present the
experimental evaluation. Finally, in Sect.~\ref{sec:conclusions}, we
conclude and discuss future work.

\section{Related Work}
\label{sec:related}
The ROSACE case study has been originally studied
in~\cite{pagetti2014rosace}. The authors model the system in
Matlab/Simulink and use simulations to evaluate it with respect to the
real-time requirements of stability, performance, and
robustness. Moreover, the authors show how to design the flight
controller as a multi-threaded software, how to do the coding and how
to validate it on a real target. However, they do not consider
faults. In this work, we assess the impact of sensor faults (such as
noise and ramp-down) and we formally validate an extended version of
the system with a simple FDIR mechanism based on a TMR architecture.

Several studies build upon ROSACE as a benchmark, focusing on
different aspects of its validation or deployment, such as code
generation~\cite{Colaco2018Scade6,Fort2019CodeGeneration}, scheduling
and execution on a multi-core
platform~\cite{PMEPST16,Trub2017PartitionedMCS,Schuh2020PredictableExecutionModels},
interference and response time
analysis~\cite{rihani:hal-01406145,Meunier2022CorrectnessEfficiency}.
The most relevant works are~\cite{R-ROSACE,R-ROSACE-2}, which describe
R-ROSACE, a redundant version of the system with replicated flight
controllers, and different variants of the model (discrete,
continuous, hybrid). Our work is orthogonal, in that it deals with
sensor faults and redundancy. Moreover, we focus on simulation-based
verification with formal guarantees for the coverage
criteria. \cite{AADL-FMI} discusses co-simulation of AADL models
(representing the controllers) and FMUs (Functional Mock-up Units) for
the physical components, and exemplified on ROSACE. As a difference,
we co-simulate the FMUs for the physical components with the embedded
SW components (FDIR logic) written in SDL. Finally, \cite{DDP15}
focuses on synthesizing hardened (more robust) architectures using
safety patterns and SMT; for ROSACE, they consider sensor failures and
failures of the execution platform, and their analysis results in a
hardened architecture with TMR patterns.
%

TASTE~\cite{TASTE2} has been used for designing embedded, real-time
systems, notably in the space domain, see,
e.g.,~\cite{ADE,MOSAR,FBK-energy}, whereas COMPASTA has been used to
analyze the motor-kit case study, a model of a DC motor drive
system~\cite{CEAS24}. Recently, COMPASTA has been extended in the
AIFDIR project~\cite{ISoLA24} to enable modeling of AI-based
components, notably for FDIR, and evaluated on an extension of the
motor-kit with AI components for anomaly detection. COMPASTA includes
the nuXmv~\cite{cavada2014nuxmv} model checker and the
xSAP~\cite{xsap-journal} safety analysis platform as back-ends. xSAP
has been used to model and analyze the TMG (Triple Modular
Generator)~\cite{xsap-journal}, a redundant, high-integrity power
distribution system.

\section{The Rosace Case Study}
\label{sec:rosace}
ROSACE (Research Open-Source Avionics and Control) is an open-source
avionics benchmark introduced in~\cite{pagetti2014rosace} to study
the full design chain of a real-time embedded flight controller, from
Matlab/Simulink modeling to multi/many-core implementation.
The system is a \emph{longitudinal flight controller} used to control different flight
parameters: vertical speed ($V_z$), true airspeed ($V_a$), and altitude ($h$) around
a nominal operating point of $h = 10\,000\,\mathrm{m}$ and $V_a = 230\,m/s$,
corresponding to a typical cruise condition.
The controller is designed to meet a set of time-domain performance properties on the
transient and steady-state response to step inputs: settling time, overshoot, rise time,
and steady-state error.

\begin{figure}[t]
	\centering
	\includegraphics[width=0.95\linewidth, scale=1.3]{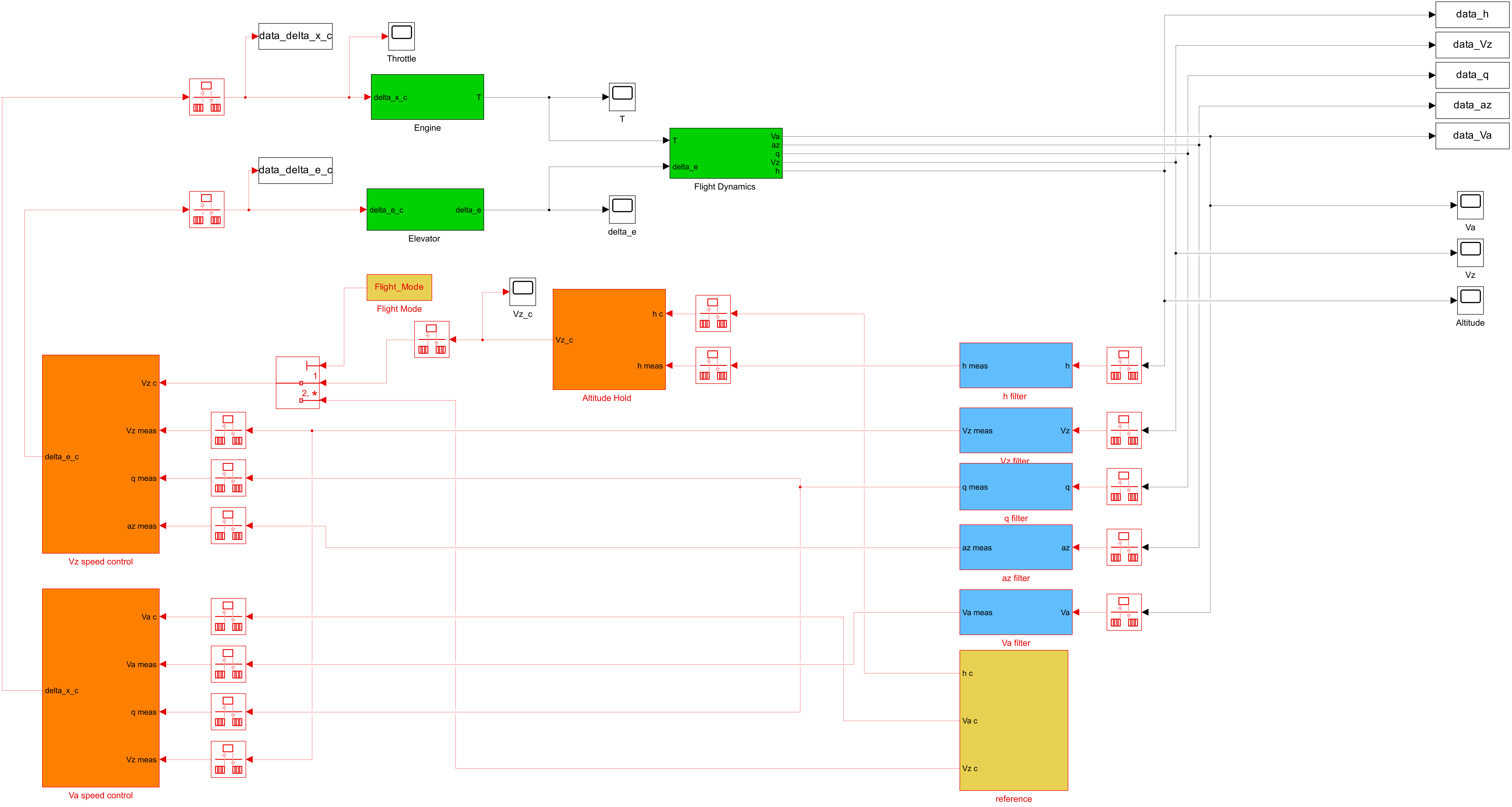}
	\caption{ROSACE Simulink architecture. Green: physical plant (flight dynamics, engine, elevator); orange: controllers; blue: sensors and filters; yellow: reference signals; red: rate-transition buffers.}
	\label{fig:rosace-arch}
\end{figure}

\paragraph{Architecture.}
The ROSACE system is composed of three interacting sub-controllers and a set of sensor
filters, each running at its own periodic rate~\cite{pagetti2014rosace}:
\begin{itemize}
	\item \textbf{Altitude hold controller} (50\,Hz): generates a $V_z$ reference from the
	altitude error $h_{\mathit{ref}} - h$; the maximum commanded $V_z$ is
	$2.5\,m/s$.
	\item \textbf{Vertical speed controller} (50\,Hz): generates the elevator deflection
	command $\delta_e$ from the $V_z$ error.
	\item \textbf{Airspeed controller} (100\,Hz): generates the throttle command $\delta_{xc}$
	from the $V_a$ error.
\end{itemize}
Sensor filters run at 100\,Hz to feed the controllers; the environment simulation
(aircraft dynamics, engine, elevator) runs at 200\,Hz to approximate the continuous-time
plant.
The two controlled outputs $V_z$ and $V_a$ are designed to be nearly decoupled: a step in
$V_a$ leaves $V_z$ almost invariant, and vice versa~\cite{pagetti2014rosace}.
Figure~\ref{fig:rosace-arch} shows the architecture of the ROSACE system, as modeled in
Simulink. The physical plant is shown in green, the controllers in orange, the sensors and
filters in blue, the references in yellow, and the rate-transition buffers in red.

\paragraph{Evaluation scenarios.}
The ROSACE benchmark considers the longitudinal motion of a medium-range civil aircraft in
the \emph{en-route} phase, which comprises two subphases~\cite{pagetti2014rosace}: the
\emph{cruise} subphase, where the autopilot maintains a constant altitude while the
autothrottle holds the airspeed $V_a$; and the \emph{change of cruise level} subphase,
where the autopilot commands a constant vertical speed $V_z$ (rate of climb) to capture a
new flight level (e.g.\ FL300\,$\to$\,FL320).
Two evaluation scenarios are derived from these subphases, each exercising the system with
a step command on one controlled variable while the others are held at trim.

\textbf{Scenario~1 change of cruise level (Sc.~1).}
The $V_z$ reference is stepped directly from $0\,m/s$ to
$-2.5\,m/s$, bypassing the altitude hold controller and commanding a climb
towards a new flight level (negative $V_z$ denotes upward motion in the downward-positive
z-axis convention used by the model).
The value $2.5\,m/s$ corresponds to the maximum vertical speed that the
altitude hold controller is allowed to command during a step climb.
The vertical speed controller is the sole primary actor, generating the elevator deflection
$\delta_e$; $V_a$ remains close to its trim value owing to the near-decoupling of the two
loops.


\textbf{Scenario~2 cruise speed change (Sc.~2).}
The true airspeed reference $V_a$ steps from its trim value of $230\,m/s$
to $235\,m/s$, commanding a small increase in cruise speed.
The airspeed controller is the sole primary actor, generating the throttle command
$\delta_{xc}$; $V_z$ remains close to zero owing to the near-decoupling of the two loops.

Both scenarios start from the same trim condition ($h =
10\,000\,\mathrm{m}$, $V_a = 230\,m/s$) and are used to assess the
controller's transient and steady-state behaviour under nominal
conditions and under faults, with and without TMR.

\section{Extending the Simulink Model}
\label{sec:modeling-simulink}
We model and analyze ROSACE and its extensions using two different
paradigms, namely we use Simulink as in~\cite{pagetti2014rosace} to
perform Simulation-Based Safety Analysis (SBSA), and
COMPASTA~\cite{CEAS24} to perform Model-Based Safety Analysis (MBSA).
The two approaches differ in some modeling choices, in the way faults
are defined/injected, and in the verification workflow.

\begin{figure}[t]
	\centering \resizebox{\textwidth}{!}{ \tikzset{
	nominal/.style={
		draw=green!30!black, fill=green!55!black, text=white,
		rounded corners=2.5pt, minimum width=2.8cm, minimum height=1.5cm,
		align=center, line width=1.2pt, font=\sffamily\small
	},
	extended/.style={
		draw=orange!50!black, fill=orange!80!yellow, text=black,
		rounded corners=2.5pt, minimum width=2.6cm, minimum height=1.3cm,
		align=center, line width=1.2pt, font=\sffamily\small
	},
	libblock/.style={
		draw=red!40!black, fill=red!55!black, text=white,
		rounded corners=2.5pt, minimum width=1.7cm, minimum height=0.75cm,
		align=center, line width=0.9pt, font=\sffamily\small
	},
	injector/.style={
		draw=blue!55!black, fill=blue!12!cyan!20, text=black,
		rounded corners=2.5pt,
		minimum width=1.2cm, minimum height=1.3cm,
		align=center, line width=0.9pt, font=\sffamily\scriptsize
	},
	faultchip/.style={
		draw=magenta!70!black, fill=magenta!60!red!80, text=white,
		rounded corners=2pt, minimum width=1.7cm, minimum height=0.65cm,
		align=center, line width=0.8pt, font=\sffamily\scriptsize
	},
	process/.style={
		draw=orange!50!black, fill=orange!5, rounded corners=5pt,
		line width=1pt, dashed, inner sep=10pt
	},
	arr/.style={
		-{Stealth[length=6pt, width=4pt]}, line width=1.3pt
	},
	fi/.style={
		draw=teal!60!black, fill=teal!18!white, text=black,
		rounded corners=3pt, minimum width=2.8cm, minimum height=1.5cm,
		align=center, line width=1.2pt, font=\sffamily\small
	},
	compasta/.style={
		draw=violet!70!black, fill=violet!12!white, text=black,
		rounded corners=3pt, minimum width=2.8cm, minimum height=1.5cm,
		align=center, line width=1.2pt, font=\sffamily\small
	},
	compasta-proc/.style={
		draw=violet!60!black, fill=violet!5, rounded corners=5pt,
		line width=1pt, dashed, inner sep=10pt
	},
	export/.style={
		-{Stealth[length=6pt, width=4pt]}, line width=1.1pt,
		draw=teal!55!black
	},
}

\begin{tikzpicture}[font=\sffamily]
	
	
	\node[nominal] (nominal) at (0, -3.6) {Nominal\\Simulink\\Model};
	
	\node[nominal] (tmr) at (0, 3.6) {TMR\\Simulink\\Model};
	
	\node[extended]  (etmr)   at (4.5,  3.6)   {TMR\\Simulink\\Model};
	\node[injector]  (inj1)   at (6.85, 3.6)   {Fault\\Definition};
	\node[faultchip] (ramp1)  at (8.82, 3.925) {Rampdown};
	\node[faultchip] (noise1) at (8.82, 3.215) {Noise};

	\node[extended]  (enom)   at (4.5,  -3.6)  {Nominal\\Simulink};
	\node[injector]  (inj2)   at (6.85, -3.6)  {Fault\\Definition};
	\node[faultchip] (ramp2)  at (8.82, -3.275){Rampdown};
	\node[faultchip] (noise2) at (8.82, -3.985){Noise};
	
	\begin{scope}[on background layer]
		\node[process, fit=(etmr)(inj1)(ramp1)(noise1)] (tmr-proc) {};
		\node[process, fit=(enom)(inj2)(ramp2)(noise2)] (nom-proc) {};
	\end{scope}
	
	\node[above=4pt of tmr-proc, font=\sffamily\small\bfseries, text=orange!60!black]
	{Extended TMR Model};
	\node[above=4pt of nom-proc, font=\sffamily\small\bfseries, text=orange!60!black]
	{Extended Nominal Model};
	
	\draw[arr] (nominal.north) -- (tmr.south);
	\draw[arr] (nominal.east) -- (nom-proc.west);
	\draw[arr] (tmr.east)      -- (tmr-proc.west);

	\draw[line width=0.85pt] (inj1.east) -- ++(0.15,0) |- (ramp1.west);
	\draw[line width=0.85pt] (inj1.east) -- ++(0.15,0) |- (noise1.west);
	\draw[line width=0.85pt] (inj2.east) -- ++(0.15,0) |- (ramp2.west);
	\draw[line width=0.85pt] (inj2.east) -- ++(0.15,0) |- (noise2.west);
	\draw[line width=0.85pt] (inj1.west) -- (etmr.east);
	\draw[line width=0.85pt] (inj2.west) -- (enom.east);
	
	
	\node[compasta]  (comp-nom-tmr)  at (13.2,  3.9)   {TMR Model};
	
	\node[compasta]  (comp-nom-single)  at (13.2,  -3.9)   {Nominal Model};
	
	\node[fi]  (fault-injector)  at (17.9,  -0.5)   {Fault\\Injector};
	\node[compasta]  (comp-tmr)  at (21.2,  3.9)   {Extended TMR Model};

	\node[compasta]  (comp-nom)  at (21.2, -3.9)   {Extended Nominal Model};

	\node[libblock] (libnoise) at (13.2,  0) {Noise};
	\node[libblock] (libramp)  at (13.2, -1) {Rampdown};

	\begin{scope}[on background layer]
			\node[draw=red!35!black, fill=pink!22, rounded corners=8pt,
		line width=1pt, inner sep=12pt,
		fit=(libnoise)(libramp)] (lib) {};
	\end{scope}
	
	\node[above=1pt of lib, font=\sffamily\small\bfseries, text=red!40!black]
	{Faults Library};

	\draw[arr, green] (comp-nom-tmr.east) -| ([xshift=-5mm]fault-injector.north);
	\draw[arr, green] ([yshift=2.5mm]fault-injector.east) -| (comp-tmr.south);
	\draw[arr, blue] (comp-nom-single.east) -| ([xshift=-5mm]fault-injector.south);
	\draw[arr, blue] ([yshift=-2.5mm]fault-injector.east) -| (comp-nom.north);
	\draw[arr, red] (lib.east) -- (fault-injector.west);
	
	\begin{scope}[on background layer]
		\node[draw=black!55, fill=none, dashed, rounded corners=8pt,
		line width=1.4pt, inner sep=18pt,
		fit=(nominal)(tmr)(tmr-proc)(nom-proc)] (simulink-side) {};
		\node[draw=violet!55!black, fill=none, dashed, rounded corners=8pt,
		line width=1.4pt, inner sep=18pt,
		fit=(comp-nom-single)(comp-nom-tmr)(comp-tmr)(comp-nom)(lib)] (compasta-side) {};
	\end{scope}

	\node[above=6pt of simulink-side, font=\sffamily\normalsize\bfseries,
	text=black!65] {Simulink Side};
	\node[above=6pt of compasta-side, font=\sffamily\normalsize\bfseries,
	text=violet!70!black] {Compasta Side};

\end{tikzpicture}
	} \caption{\emph{Left}: Simulink-based models (nominal vs.\
	TMR); each extended model embeds its own Fault Definition
	block (a custom Simulink library block) that applies Rampdown
	and Noise faults directly to sensor outputs.  \emph{Right}:
	COMPASTA-based models, with nominal and TMR models extended
	via a shared Faults Library (Noise and Rampdown) fed through
	the COMPASTA Fault Injector
	block.}  \label{fig:modeling-overview}
\end{figure}
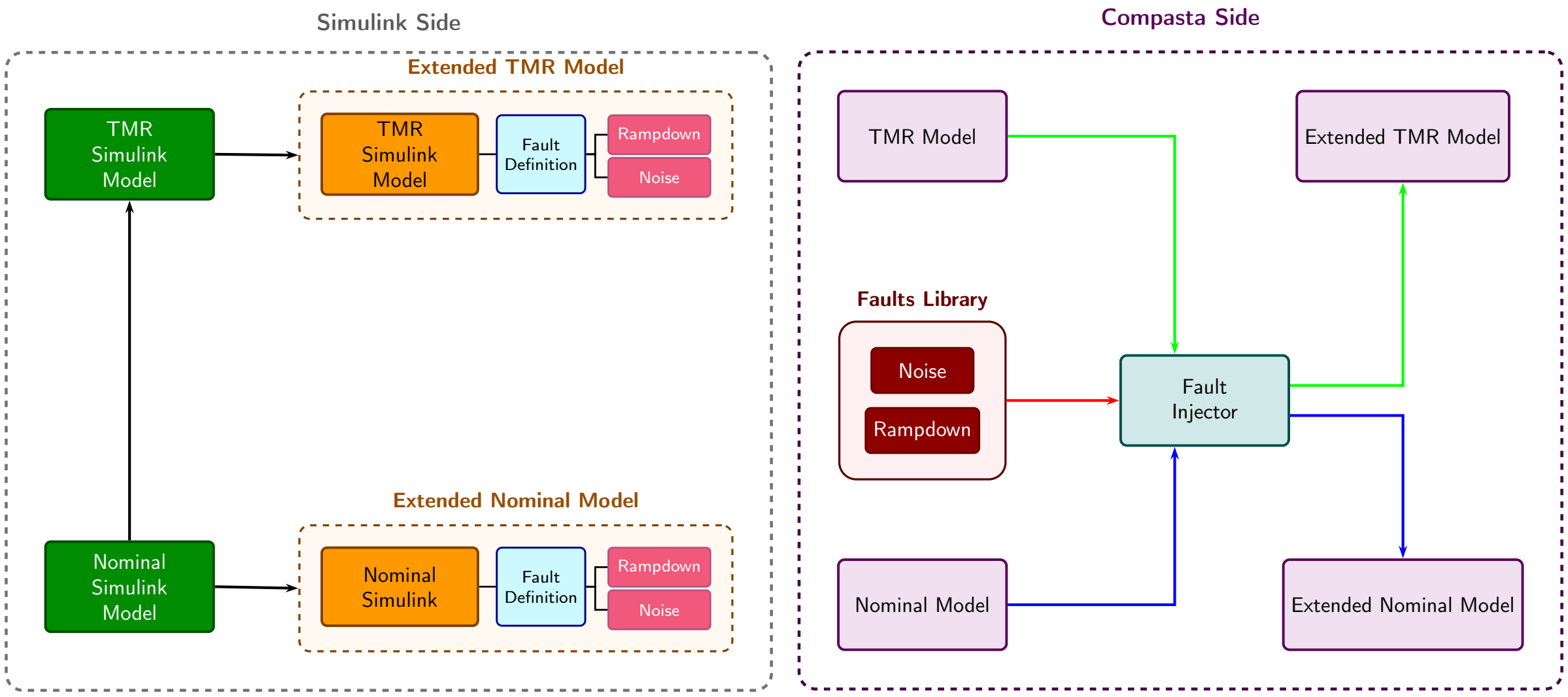

In this section, we first present the Simulink models: the original
(nominal model) and the TMR architecture, with and without faults
(applied to sensor outputs). The resulting four models are depicted in
Fig.~\ref{fig:modeling-overview} (left). Faults are modeled using
custom or standard Simulink blocks, and their activation is controlled
by Boolean signals.

\paragraph{Nominal Simulink model.}
The starting point is the non-linear Simulink model of the ROSACE
flight controller introduced in Sect.~\ref{sec:rosace}. This model
is used as the reference architecture of the system, and the basis for
the other three models.

\paragraph{TMR Simulink model.}
The TMR architecture is obtained from the nominal model by replicating
each sensor channel.  Each of the three replicas measures the same
physical quantity and feeds its output to a majority voter, whose
output replaces the single-channel measurement in the rest of the
control loop. Crucially, this architectural extension is
\emph{independent} of fault injection: the TMR model is a fault-free,
structurally redundant variant of the nominal model, and the two
architectures constitute the primary object of comparison throughout
the paper.

\paragraph{Fault Definition.}
To support fault injection, we introduce a reusable Simulink custom
subsystem of fault dynamics, which can be applied to any sensor
output.\footnote{Matlab/Simulink provides the fault analyzer module,
which can be used to inject fault behavior into a given Simulink
signal, however it is not supported by the FMU model generation.}
Let $m$ be the actual value of a measured physical quantity, $M$ the
function that computes the sensor output and $k \in \{1,2,3\}$ index
the three replicas of a sensor in the TMR architecture ($k=1$ in the
nominal one). Each faulty sensor $k$ produces an output
$
s_k = M(m,\delta_k^{\mathit{eff}})
$
where $\delta_k^{\mathit{eff}}$ is determined by the fault type and the activation pattern defined below. 

\smallskip\noindent\textit{Additive noise.}
Sensor $k$ is perturbed by an independent zero-mean white noise:

\[
\delta_k^{noise} \quad \text{with} \quad \mathbb{E}[\delta_k^{noise}] = 0, \quad \mathbb{E}[\delta_k^{noise}\delta_j^{noise}] = 
\begin{cases} 
	\sigma^2 & \text{for } k = j \\ 
	0 & \text{for } k \neq j 
\end{cases}
\]

\smallskip\noindent\textit{Ramp-down fault.}
Sensor $k$ is affected over the time by a signed, magnitude deviation:
$
t\cdot\delta^{ramp}; \quad\delta^{ramp}> 0,
$  

\paragraph{Fault Activation Pattern.}
Both fault types share the same activation logic.  A binary mask $p =
(p_1, p_2, p_3)$ with $p_k \in \{0,1\}$ determines which sensors are
affected by the fault, with $p_k = 1$ indicating that sensor $k$ is
faulty and $p_k = 0$ that it is healthy.
The effective perturbation ($\delta_k^{\mathit{eff}}$) applied to
sensor $k$ is defined as: $
\delta_k^{\mathit{eff}} = p_k \cdot \delta_k^{noise|ramp}.
$
Masks are grouped by fault multiplicity $|p| = \sum_k p_k$ (from 0
to 3, corresponding to absence of faults, single, double, or triple
faults). The sensor output is then defined as follows, where the
absolute lower limit 0 enforces a realistic physical saturation
profile during the sensor degradation.
%
\begin{equation}
	M(m, \delta_k^{eff}) \rightarrow 
	\begin{cases}
		M(m, 0) = m\\
		M(m, \delta^{noise}) = m + \delta^{noise}\\
		M(m, \delta^{ramp}) = \text{sign}(m) \cdot \max( |m| - t\cdot\delta^{ramp}, \; 0
)
	\end{cases}
\end{equation}

\paragraph{Extended Models.}
The fault blocks from the fault dynamic library are imported into 
the nominal and TMR models, with the same injection pattern, producing the 
\emph{Extended Nominal Model} and the \emph{Extended TMR Model}, respectively.
Fault analysis is performed by simulating the extended models under
different fault injection patterns; the resulting traces are then
evaluated against the system properties as described in
Sect.~\ref{sec:analysis}.

\subsection{Modeling Voter scheme}
\label{sec:modeling-voter}

The majority voter is the functional element that implements
redundancy management in the TMR architecture. It aggregates the three
sensor readings and selects the majority value to be forwarded to the
controller.

The voting logic is common to both modeling paradigms, Simulink
and COMPASTA, but with different modeling choices. In the Simulink
model, the voter is implemented using basic Simulink blocks (e.g.,
comparators, logical operators).  In the COMPASTA model, the voter is
implemented as a reactive process specified in SDL.

\paragraph{Voter specification.}
Let $s_k$ denote the output of replica $k \in \{1,2,3\}$ after fault injection,
as defined in Section~\ref{sec:modeling-simulink}, and
$\epsilon > 0$ a threshold parameter.
The voter $V: \mathbb{R}^3 \to \mathbb{R}$ is defined as follows:
\[
\hat{m} \;=\; V(s_1, s_2, s_3) \;=\;
\begin{cases}
	\dfrac{1}{3}\displaystyle\sum_{k=1}^{3}s_k
	& \text{if } \lvert s_k - s_j\rvert \leq \epsilon
	\text{ or }
	\lvert s_k - s_j\rvert > \epsilon
	\text{ for all } k \neq j, \\[14pt]
	\dfrac{s_k + s_j}{2}
	& \text{if } \exists!\,(k,j),\ k < j,\text{ such that }
	\lvert s_k - s_j\rvert \leq \epsilon.
\end{cases}
\]
The first branch covers full consensus and full disagreement: in both
cases the arithmetic mean is used, either to aggregate all agreeing
readings or as a graceful degradation fallback. The second branch masks a single faulty replica by
outputting the midpoint of the uniquely agreeing pair.

\section{Modeling in COMPASTA}
\label{sec:modeling-compasta}
In this section, we first discuss the ROSACE system requirements and
their formalization in Sect.~\ref{sec:formal-requirements}. Then, we
discuss the formal models in Sect.~\ref{sec:formal-models}.

\subsection{Formal Requirements}
\label{sec:formal-requirements}
The ROSACE benchmark comes with four control-theoretic requirements
(P1--P4). We first describe them in natural language as
in~\cite{pagetti2014rosace}. A key advantage of MBSA is that the
system requirements can be expressed formally in temporal logic,
enabling rigorous, automated verification. Specifically, we formalize
each requirement in linear temporal logic (LTL) and its
extensions~\cite{alur1994tptl,alur1993real,koymans1990specifying}.


A key challenge in formalizing these requirements is that the original control-theoretic definitions
express tolerances relative to the \emph{steady-state value} of the output, that is
only known after the simulation ends and therefore cannot be used as a bound during formal
verification or on-line property monitoring.
To overcome this, we express all tolerances relative to the \emph{step amplitude}
$\Delta_F$, which is fully determined at $t = 0$ from the step command and the initial output:
$
\Delta_F \;=\; \bigl|\mathit{target} - \mathit{output}(0)\bigr|
$.

This substitution is conservative: since the true steady-state value lies within
a known neighborhood of \emph{target} (guaranteed by the steady-state error requirement),
using \emph{target} in place of the true steady-state only tightens the bounds,
making all derived requirements more stringent.
We write $\mathit{output}(t)$ for the controlled output at time $t$ and $\mathit{target}$ for
the commanded reference value.

\paragraph{\textbf{P1: Settling time.}}
\textit{After a step command, the output must reach and permanently stay within a tolerance band of $\delta_1\cdot\Delta_F$ around the target, no later than $\tau_1$ time units after the command.}
Formally, there exists a time in $[0, \tau_1]$ from which the output deviation never exceeds again the tolerance band:
\begin{equation}\label{eq:p1}
	\Diamond_{[0,\,\tau_1]}\;\left(\square\!\left(\bigl|\mathit{output} - \mathit{target}\bigr|\leq\delta_1\cdot\Delta_F\right)\right).
\end{equation}

\paragraph{\textbf{P2: Overshoot.}}
\textit{At all times, the absolute output must not exceed the absolute target value more than a fraction $\delta_2$ of the target magnitude.}
Formally, the output must always be below the overshoot threshold:
\begin{equation}\label{eq:p2}
	\square\!\left(\mathit{target} - output \leq\delta_2\cdot\Delta_F\right).
\end{equation}

\paragraph{\textbf{P3: Rise time.}}
\textit{The output must react promptly to the command: once it has
	crossed a first response threshold $\delta_{3_1}\cdot\Delta_F$ away
	from the initial value $\mathit{output}_0$, it must reach a higher
	threshold $\delta_{3_2}\cdot\Delta_F$ within $\tau_3$ additional time
	units.}

The thresholds $\delta_{3_1} = 10\%$ and $\delta_{3_2} = 90\%$
correspond to the standard 10\%--90\% rise-time
definition~\cite{pagetti2014rosace}, expressed here relative to
$\Delta_F$ rather than to the steady-state value (see the discussion
at the start of this section).

Formally, at every instant $t_{\delta_{3_1}}$ where the output crosses
the first threshold, it must reach the second threshold within $\tau_3$
time units:
\begin{equation}\label{eq:p3}
	\Box\Bigl(
	\bigl|\mathit{output}_{t_{\delta_{3_1}}} - \mathit{output}_{0}\bigr|
	\geq \delta_{3_1}\cdot\Delta_F \implies{}
	\Diamond_{[0,\,\tau_3]}\bigl(
	\bigl|\mathit{output}_{t} - \mathit{output}_{0}\bigr|
	\geq \delta_{3_2}\cdot\Delta_F
	\bigr)
	\Bigr)
\end{equation}

\paragraph{\textbf{P4: Steady-state error.}}
\textit{The output error with respect to the target must remain bounded
within $\delta_4\cdot\Delta_F$ after the settling time $\tau_1$.}

Formally, at some point ($\tau_4$) after the settling time  the deviation from the
target must never exceed the tolerance band throughout the remaining time:

\begin{equation}\label{eq:p4}
	\Box_{[\tau_4,\,T]} \left(\bigl|\mathit{output}-\mathit{target}\bigr|\leq\delta_4\cdot\Delta_F\right).\quad \text{with } \tau_4 \geq \tau_1
\end{equation}
\medskip
Table~\ref{tab:props} summarizes the parameter values of the four
requirements for the two main scenarios.

\begin{table}[H]
	\centering
	\caption{ROSACE requirements: $\tau$ and $\delta$ values per scenario.~\cite{pagetti2014rosace}}
	\label{tab:props}
	\begin{tabular}{lcccc}
		\toprule
		& \multicolumn{2}{c}{\textbf{Sc.~1 ($V_z = -2.5\,\mathrm{m/\,s}$)}}
		& \multicolumn{2}{c}{\textbf{Sc.~2 ($V_a = 235\,\mathrm{m/\,s}$)}} \\
		\cmidrule(lr){2-3}\cmidrule(lr){4-5}
		\textbf{Property}
		& $\tau$               & $\delta$
		& $\tau$               & $\delta$ \\
		\midrule
		P1: Settling time      & $\leq 10\,\mathrm{s}$ & $\leq 5\%$  & $\leq 20\,\mathrm{s}$ & $\leq 5\%$  \\
		P2: Overshoot          & ---                   & $\leq 10\%$ & ---                   & $\leq 10\%$ \\
		P3: Rise time          & $\leq 6\,\mathrm{s}$  & ---         & $\leq 12\,\mathrm{s}$ & ---         \\
		P4: Steady-state error & ---                   & $\leq 5\%$  & ---                   & $\leq 5\%$  \\
		\bottomrule
	\end{tabular}
\end{table}

\subsection{Formal Models.}
\label{sec:formal-models}
The COMPASTA formal models are specified in the Slim (a variant of
AADL) and SDL languages~\cite{CEAS24} . They provide an abstract
representation of the system used for model-based reasoning, whereas
formal verification is based on the co-simulation of the formal models
with the corresponding physical Simulink model exported as an FMU
(compare Section~\ref{sec:analysis}). COMPASTA models are summarized
in Fig.~\ref{fig:modeling-overview} (right).
Both the nominal and TMR architectures are represented in Slim.
Namely, each Simulink subsystem becomes a Slim component whose ports
mirror the Simulink signal interfaces.  Figure~\ref{fig:compasta-tmr}
shows the COMPASTA TMR architecture. It contains the following blocks:
\begin{figure}[t]
	\centering
	\includegraphics[width=0.95\linewidth]{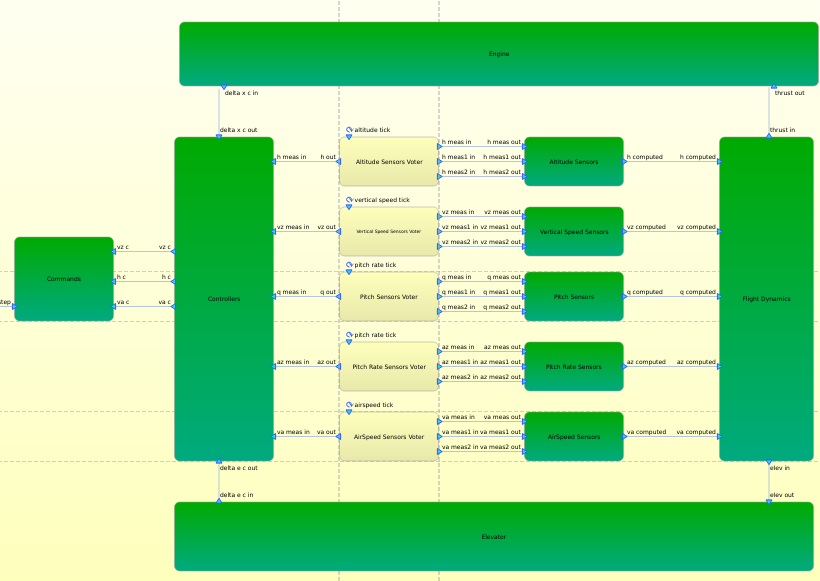}
	\caption{TMR architecture in the COMPASTA
          environment. Dark-green blocks (\emph{flight dynamics},
          \emph{engine}, \emph{elevator}, \emph{controllers}, sensor
          groups) are direct counterparts of the nominal Simulink
          subsystems. Yellow Voter blocks implement the majority voter
          for each sensor channel (see
          Sect.~\ref{sec:modeling-voter}).}
	\label{fig:compasta-tmr}
\end{figure}
\begin{itemize}
	\item \textbf{Plant and actuation blocks} (dark green)
          \emph{flight dynamics}, \emph{engine}, and \emph{elevator}
          correspond directly to the nonlinear Simulink plant and
          actuator blocks of Sect.~\ref{sec:rosace}.  The \emph{flight
          dynamics} block is linearized within the Slim model, a
          necessary step to make the continuous-time plant amenable to
          symbolic analysis; all other blocks are represented
          exactly.\footnote{Linearization is the sole source of
          approximation in the COMPASTA approach. It affects the
          generation of the abstract test scenarios (via model
          checking), but not the concrete execution traces
          (corresponding to the test scenarios), which are obtained
          via co-simulation of the non-linear Simulink models. Hence,
          it does not affect the results of the safety analysis.}
	
	\item \textbf{Controllers block} (dark green)
	The \emph{controllers} component mirrors the three Simulink
	sub-controllers (altitude hold, vertical speed, and airspeed
	controllers) as a single Slim composite, preserving the
	same periodic rates and signal interfaces.
	
	\item \textbf{Sensor replica blocks} (dark green)
	Five sensor groups correspond to the five measured physical
	quantities: altitude ($h$), vertical speed ($V_z$), pitch rate
	($q$), vertical acceleration ($a_z$), and airspeed ($V_a$).
	Each group contains three replicas (ports \texttt{meas},
	\texttt{meas1}, \texttt{meas2}), matching the TMR replication
	introduced in the Simulink TMR model
	(Sect.~\ref{sec:modeling-simulink}).
	
	\item \textbf{FDIR blocks} (yellow)
	Each sensor channel is paired with a dedicated
	\emph{$\langle$quantity$\rangle$ sensors Voter} block
	(\emph{altitude sensors Voter}, \emph{vertical speed sensors Voter},
	\emph{pitch sensors Voter}, \emph{pitch rate sensors Voter},
	\emph{airspeed sensors Voter}).
	These components implement the majority
	voter of Sect.~\ref{sec:modeling-voter}: they receive the three
	replica measurements, select the majority value, and forward the
	consolidated signal (\texttt{h computed}, \texttt{vz computed},
	etc.) to the \emph{controllers} block.
	They are specified in SDL as reactive processes.
	
\end{itemize}

\noindent
The nominal COMPASTA model shares the same architecture, with the
\emph{flight dynamics}, \emph{engine}, \emph{elevator}, and
\emph{controllers} blocks retained as-is, but without the sensor
replica channels and the FDIR voter blocks.

\paragraph{Fault specification.}
Faults are specified in a declarative manner, and attached to
individual sensor components. The COMPASTA fault injector takes care
of automatically producing the extended models. In the TMR model, each
of the three replicas per sensor family carries its own fault
definition; fault activation is governed by the same Boolean signal
$p_k \in \{0,1\}$ introduced in Sect.~\ref{sec:modeling-simulink}.

\section{Formal Analysis}
\label{sec:analysis}
We describe how the two modeling paradigms introduced in
Sect.~\ref{sec:modeling-simulink} and~\ref{sec:modeling-compasta} are
used to perform the analyses.  Both target the same requirements
P1--P4 (Sect.~\ref{sec:formal-requirements}) and the same fault
injection scenarios, but they differ significantly in how properties
are checked and in the guarantees they provide.
In the \emph{Simulation-Based Safety Analysis} (SBSA) workflow, the
Simulink models are simulated, on sampled execution traces, under the
fault activation patterns defined in
Sect.~\ref{sec:modeling-simulink}; the resulting traces are then
evaluated against the requirements by dedicated monitoring
instrumentation.  In the \emph{Model-Based Safety Analysis} (MBSA)
workflow, COMPASTA uses the formal model (abstracting both the
environment and the system components) to automatically generate a set
of test scenarios, which are guaranteed by construction to cover the
state space according to the chosen notion of coverage (e.g., at least
one test for each evaluation scenario, as per Sect.~\ref{sec:rosace},
with up to two faults). The test scenarios are simulated, and
properties expressed in temporal logic are formally evaluated on the
simulation traces, without manual monitor construction.

\paragraph{COMPASTA toolchain pipeline.}
\begin{figure}[t]
	\centering
	\resizebox{0.8\textwidth}{!}{%
\tikzset{
  inpblock/.style={
    draw=black!50, fill=white, rounded corners=4pt,
    minimum width=2.2cm, minimum height=0.9cm,
    align=center, line width=0.9pt, font=\sffamily\small
  },
  procblock/.style={
    draw=gray!60!black, fill=gray!24, rounded corners=10pt,
    minimum width=2.0cm, minimum height=0.82cm,
    align=center, line width=1.1pt, font=\sffamily\small
  },
  fileart/.style={
    chamfered rectangle, chamfered rectangle corners=north east,
    chamfered rectangle sep=6pt,
    draw=#1!65!black, fill=#1!20,
    minimum width=2.0cm, minimum height=1.0cm,
    align=center, line width=0.9pt, font=\sffamily\scriptsize, text=black
  },
  central/.style={
    draw=gray!65!black, fill=gray!10, rounded corners=6pt,
    minimum width=2.8cm, minimum height=6.4cm,
    align=center, line width=1.3pt, font=\sffamily\small\bfseries
  },
  icongreen/.style={
    draw=green!50!black, fill=green!22, rounded corners=3pt,
    minimum width=1.7cm, minimum height=0.75cm,
    align=center, line width=0.9pt, font=\sffamily\scriptsize
  },
  iconmag/.style={
    draw=magenta!60!black, fill=magenta!18, rounded corners=3pt,
    minimum width=1.7cm, minimum height=0.75cm,
    align=center, line width=0.9pt, font=\sffamily\scriptsize
  },
  verifblock/.style={
    draw=blue!45!black, fill=blue!12, rounded corners=4pt,
    minimum width=2.1cm, minimum height=0.9cm,
    align=center, line width=0.9pt, font=\sffamily\scriptsize, text=black
  },
  fmudetail/.style={
    draw=gray!50, fill=gray!40, rounded corners=3pt,
    minimum width=2.8cm, minimum height=2.2cm,
    align=center, line width=0.9pt, font=\sffamily\scriptsize, text=white
  },
  garr/.style={-{Stealth[length=5pt,width=3.5pt]},
               line width=1.0pt, draw=green!55!black},
  barr/.style={-{Stealth[length=5pt,width=3.5pt]},
               line width=1.0pt, draw=black!65},
  parr/.style={-{Stealth[length=5pt,width=3.5pt]},
               line width=1.0pt, draw=blue!60!purple},
}

\begin{tikzpicture}[font=\sffamily]


  \node[inpblock]          (dynmodel) at (0,    4.2)  {Dynamic\\Model};
  \node[procblock]         (fmieng)   at (3.4,  4.2)  {FMI\\Engine};
  \node[fileart=blue]      (fmu)      at (7.0,  4.2)  {FMU\\Co-Simulation\\Model};

    \node[draw=gray!55!black, fill=gray!18, rounded corners=5pt,
        minimum width=2.4cm, minimum height=1.1cm,
        align=center, line width=1.1pt, font=\sffamily\small]
        (code-gen) at (0.5, 1.2) {Code Generator};
  \node[fileart=yellow!80!orange] (jsondom)  at (3.4,  0.2)  {Scenario\\Definition};
  \node[inpblock]                 (slimsdl)  at (0.5, -1.1)  {FORMAL MODEL\\SLIM + SDL};
  \node[icongreen]                (fprop)    at (-0.5,-2.7)  {Formal\\Properties};
  \node[iconmag]                  (faultdef) at (1.5, -2.7)  {Faults\\Definition};
  \node[procblock]                (smveng)   at (3.4, -1.1)  {Translators};
  \node[fileart=cyan!80!teal]     (smvcode)  at (7.0, -1.1)  {SMV\\Code};

  \node[draw=gray!55!black, fill=green!18, rounded corners=5pt,
        minimum width=2.4cm, minimum height=1.1cm,
        align=center, line width=1.1pt, font=\sffamily\small]
        (simulator) at (10.5, 4.2) {Simulator};
  \node[draw=gray!55!black, fill=blue!18, rounded corners=5pt,
        minimum width=2.4cm, minimum height=1.1cm,
        align=center, line width=1.1pt, font=\sffamily\small]
        (modelchecker) at (10.5, -1.1) {Model\\Checker};

  \begin{scope}[on background layer]
    \node[central, fit=(simulator)(modelchecker), inner sep=12pt,
          label={[font=\sffamily\small\bfseries, text=gray!60!black]above:COMPASTA}]
          (tastesim) {};
  \end{scope}

  \node[fileart=green!85!red] (trace) at (14.3, 4.2)  {Execution\\Trace};
  \node[fileart=orange!85!red] (abstract) at (14.3, 0.5)  {Abstract\\Scenario};
  \node[fileart=cyan]          (concrete) at (14.3, 2.2) {Concrete\\Scenario};
  \node[verifblock]            (verif)    at (14.3,-1.1)  {Formal Properties\\Verification};


  \begin{scope}[on background layer]
    \node[draw=orange!55!black, fill=orange!4,
          rounded corners=6pt, line width=0.9pt, dashed,
          inner sep=10pt,
          fit=(dynmodel)(fmieng)(fmu)] (simbox) {};
    \node[draw=blue!50, fill=blue!5,
          rounded corners=6pt, line width=0.9pt, dashed,
          inner sep=14pt,
          fit=(code-gen)(jsondom)(slimsdl)(fprop)(faultdef)(smveng)(smvcode)] (tastebox) {};
  \end{scope}

  \node[anchor=north west, font=\sffamily\scriptsize\bfseries, text=orange!70!black]
    at ($(simbox.north west)+(4pt,-4pt)$)   {Matlab/Simulink};
  \node[anchor=north west, font=\sffamily\scriptsize\bfseries, text=blue!65!black]
    at ($(tastebox.north west)+(4pt,-4pt)$) {COMPASTA};

  \coordinate (sepMid) at ($(simbox.south)!0.5!(tastebox.north)$);
  \draw[dashed, gray!55, line width=1.0pt]
    (tastebox.west |- sepMid) -- (tastebox.east |- sepMid);


  \draw[garr] (dynmodel.east)  -- (fmieng.west);
  \draw[garr] (fmieng.east)    -- (fmu.west);
  \draw[garr] (fmu.east)       -- (simulator.west);

  \draw[barr] (jsondom.east)   -- (tastesim.west |- jsondom.east);

  \draw[parr] (fprop.north)    -- (fprop.north    |- slimsdl.south);
  \draw[parr] (faultdef.north) -- (faultdef.north |- slimsdl.south);
  \draw[parr] (slimsdl.east)   -- (smveng.west);
  \draw[parr] (smveng.east)    -- (smvcode.west);
  \draw[parr] (smvcode.east)   -- (modelchecker.west);
  \draw[parr] (code-gen.east)  -- ++(0.6,0) -| (simulator.south);
  \draw[parr] (slimsdl.north)   -- (code-gen.south);
  \draw[parr] (modelchecker.east) -- ++(0.6,0) |- (abstract.west);
  \draw[garr] (concrete.west)    -- ++ (0.0,0) -| ([xshift=0.5cm]simulator.south);
  \draw[garr] (abstract.north)    -- (concrete.south);
  \draw[parr] (simulator.east) -- (trace.west);
  \draw[garr] (trace.east)    -- ++(0.6,0) |-  (verif.east);
  \draw[parr] (modelchecker.east) -- ++(0.4,0) |- (verif.west);

\end{tikzpicture}
	}
	\caption{COMPASTA toolchain: integration of the formal model
          with the FMU co-simulation component.}
	\label{fig:compasta-toolchain}
\end{figure}
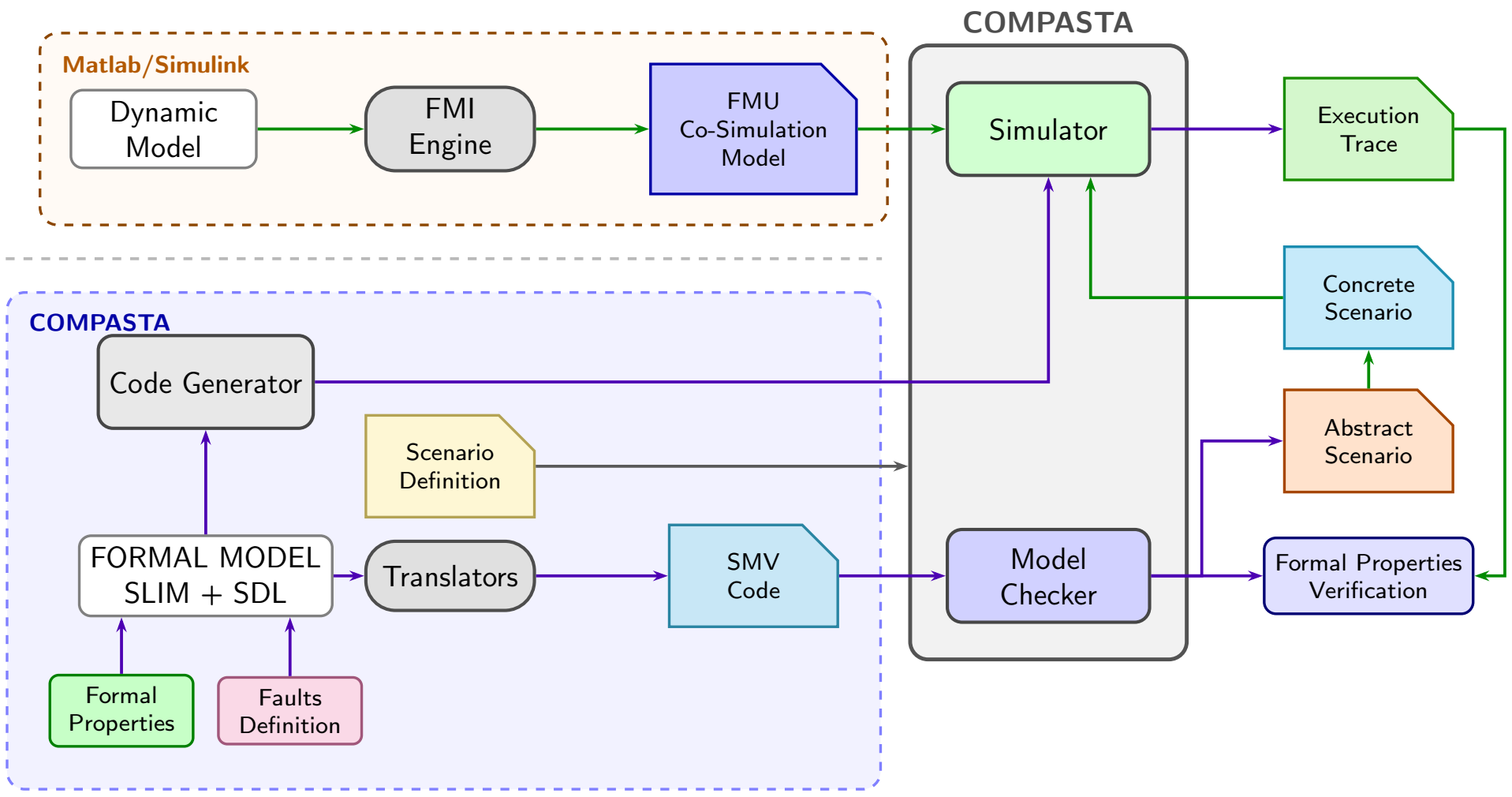
COMPASTA processes the inputs through two compilation paths
(Fig.~\ref{fig:compasta-toolchain}). On the Simulink side, the dynamic
model of the plant is compiled into a dedicated FMU (Functional
Mock-up Unit) exported via the FMI standard (Functional Mock-up
Interface).  Each model (nominal, extended, with or without TMR) has a
corresponding FMU.  In COMPASTA, the Slim/SDL model augmented with the
properties and faults definition is compiled by the SMV engine into
SMV code (the input language of the nuXmv~\cite{cavada2014nuxmv} model
checker), and used to formally generate test scenarios.

\paragraph{Coverage criteria and scenario generation.}

MBSA generates test scenarios according to explicitly defined coverage
criteria. Two complementary dimensions are considered. In \emph{Fault
multiplicity coverage}, we target all single-fault configurations
(fault in either the $V_z$ or $V_a$ channel) and all combined
two-fault configurations (simultaneous faults in both channels), as
these represent the minimal and maximal fault multiplicities relevant
to the system's nominal operation. For the TMR, the coverage criterion
is defined in terms of the number of faulty replicas per sensor group:
$P_{[0]}$ (no fault), $P_{[1]}$ (one faulty replica), $P_{[2]}$ (two
faulty replicas), and $P_{[3]}$ (all three replicas faulty). This
covers the entire spectrum of multiplicities up to the group size,
enabling verification of the voter's fault-masking capability up to
the design assumption of at most $N-1$ faulty replicas. In \emph{Fault
parameter space coverage}, for additive noise faults, the parameter
space is covered by discretizing the SNR range (Signal-to-Noise Ratio)
into five logarithmically spaced levels (35, 25, 15, 5, and
$\sim\!0\,\mathrm{dB}$), which spans from nominal operation to the
noise floor. For ramp-down faults, two rates per channel are
considered: a slow rate (gradual degradation) and a fast rate (abrupt
failure approaching saturation within the simulation horizon).  This
yields a total of $2 \times 2 = 4$ parameter combinations for
ramp-down faults.
The concretization phase maps each abstract scenario to a concrete
simulation trace by replacing abstract parameter values with their
concrete signal equivalents in the FMU model.
\begin{figure}[t]
	\centering
	\input{chapters/properties-pattern}
	\caption{SLIM property pattern declarations for the vertical
          speed scenario.}
	\label{fig:properties-pattern}
\end{figure}
\paragraph{VIVAS verification and validation workflow.}
The COMPASTA verification and validation flow (VIVAS)~\cite{ISoLA24},
consists of the following phases. The nuXmv~\cite{cavada2014nuxmv}
model checker is used to generate a set of abstract scenarios
(symbolic traces), from the formal model, translated into SMV, and a
specification of the scenarios to be covered.  These scenarios are
concretized into \emph{concrete scenarios} (amenable to simulation) by
mapping abstract parameters into the corresponding signals in the
FMU. The COMPASTA simulator is used to orchestrate the co-simulation
of the SW components (corresponding to deployed, executable versions
of the SDL components) with the FMU corresponding to the plant, step
by step, under the requested conditions specified in each
scenario. The original non-linear plant is co-simulated with the rest
of the system, preserving the original dynamics as in the
corresponding SBSA runs. The output is an execution trace of the
system, containing all the necessary details to evaluate the
properties.
\paragraph{Property evaluation in Simulink (SBSA).}
Requirements P1--P4 are evaluated against simulation traces by two
alternative instruments, each offering a different trade-off.  The
Breach toolbox~\cite{donze2010breach} encodes P1--P4 in Signal
Temporal Logic (STL) and checks them against recorded simulation
traces in an offline pass, producing a quantitative robustness margin
and a Boolean satisfiability verdict per property.  Alternatively,
dedicated \emph{monitor automata}, as in~\cite{pagetti2014rosace},
i.e., a manually constructed Simulink subsystem per property, observe
the controlled outputs in real time during simulation and emit a
binary satisfied/violated signal at each integration step.  In both
cases, only the traces actually generated by the simulator are
checked, with no guarantee of coverage over the full state space.
Moreover, the manual encoding effort required by each instrument (STL
formula or Simulink subsystem) is avoided entirely using MBSA.
\paragraph{Property evaluation in COMPASTA (MBSA).}
COMPASTA takes a fundamentally different approach.  Requirements
P1--P4 are expressed natively in Linear Temporal Logic (LTL) as
\emph{property patterns}~\cite{COMPASS1,COMPASS2} and attached
directly to the top-level Slim system implementation (see, e.g.,
Fig.~\ref{fig:properties-pattern}). No monitor automaton needs to be
constructed manually: property evaluation is carried out using a
run-time monitor, automatically generated from the formal model and the
formal specification of the property, and evaluated on the execution
traces produced by the COMPASTA simulator.

\section{Experimental Results}
\label{sec:experimental}
%

We evaluate the ROSACE benchmark under Scenario~1 ($V_z$
vertical-speed control, Sect.~\ref{sec:rosace}) with both the SBSA
(Simulink) and MBSA (COMPASTA) approaches, assessing requirements
P1--P4 (Sect.~\ref{sec:formal-requirements}) under two fault
types additive sensor noise and ramp-down degradation on the $V_z$
and $V_a$ sensor channels.  On the \emph{nominal} model, faults are
injected at the channel level across increasing severity
(Sect.~\ref{sec:results-nominal}).  On the \emph{TMR} model, faults
are characterized by their multiplicity per replica group: $P_{[0]}$
(none), $P_{[1]}$ (one faulty replica), $P_{[2]}$ (two), and $P_{[3]}$
(all three); this analysis uses MBSA and is reported in
Sect.~\ref{sec:results-tmr}.

\subsection{Experimental Setup}
\label{sec:setup}

\paragraph{Noise.}
The baseline noise variance of each channel is fixed from its maximum
physical value $X_{max}$ and a nominal $\mathrm{SNR}_0 = 35\,\mathrm{dB}$
(standard operation at $h = 10\,000\,\mathrm{m}$), as
$\sigma_0^2 = X_{max}^2 / 10^{\,\mathrm{SNR}_0/10}$; the ceiling values
$X_{max}$ follow from the aircraft physical
parameters~\cite{pagetti2014rosace} and standard flight-performance
relations~\cite{anderson1999aircraft}.  Degraded levels are obtained as
$\sigma_k^2 = k\,\sigma_0^2$ with $k \in \{1,10,100,1000,3162\}$,
yielding SNR values of $35, 25, 15, 5, \sim\!0\,\mathrm{dB}$.  The level
is set through a workspace scale factor in SBSA and through an FMU
parameter in MBSA, updated before each run.

\paragraph{Ramp-down.}
Two ramp-down rates per channel are defined as fractions of the trim
operating value ($V_{a,eq} = 230\,\mathrm{m/\,s}$;
$\Delta_F = 2.5\,\mathrm{m/\,s}$, the commanded $V_z$ step).  The
baseline noise parameters and the ramp-down rates are summarized in
Table~\ref{tab:fault-params}.

\begin{table}[H]
  \centering
  \caption{Per-channel fault parameters for Scenario~1: baseline noise
    at $\mathrm{SNR}_0 = 35\,\mathrm{dB}$ and the two ramp-down rates.
    $X_{max}$, $\sigma_0$ and the reference value in $\mathrm{m/\,s}$;
    $\sigma_0^2$ in $\mathrm{m^2/\,s}$; $\delta^{ramp}$ in
    $\mathrm{m/\,s^{2}}$.}
  \label{tab:fault-params}
  \small
  \setlength{\tabcolsep}{4pt}
  \begin{tabular}{@{}lcccccc@{}}
    \toprule
    & \multicolumn{3}{c}{\textbf{Noise (35\,dB)}}
      & \multicolumn{3}{c}{\textbf{Ramp-down}} \\
    \cmidrule(lr){2-4}\cmidrule(lr){5-7}
    Channel & $X_{max}$ & $\sigma_0^2$ & $\sigma_0$
      & Ref.\ value & $\delta_1^{ramp}$ & $\delta_2^{ramp}$ \\
    \midrule
    $V_a$ & $280.65$ & $24.9075$ & $4.9907$ & $230$ & $0.23$   & $2.3$   \\
    $V_z$ & $7.32$   & $0.01694$ & $0.1302$ & $2.5$ & $0.0025$ & $0.025$ \\
    \bottomrule
  \end{tabular}
\end{table}


\subsection{Nominal Model Results}
\label{sec:results-nominal}
Table~\ref{tab:nominal} consolidates the results obtained on the
nominal model for Scenario~1 ($V_z$ control), under noise faults (top
block, parameter = SNR) and ramp-down faults (bottom block,
parameter = $\delta^{ramp}$), for both SBSA and MBSA.  For the noise
faults only SNR levels that change the verdict pattern are reported:
results at 5\,dB and 0\,dB are identical to those at 15\,dB and are
omitted.

\begin{table}[H]
  \centering
  \caption{Consolidated nominal-model results for Scenario~1 ($V_z$
    control), under noise and ramp-down faults.
    \textbf{Y} = property satisfied; \textbf{N} = violated.
    SBSA and MBSA agree on every cell.}
  \label{tab:nominal}
  \setlength{\tabcolsep}{4pt}
  \small
  \begin{tabular}{@{}llcccccccc@{}}
    \toprule
    & & \multicolumn{4}{c}{\textbf{SBSA}}
      & \multicolumn{4}{c}{\textbf{MBSA}} \\
    \cmidrule(lr){3-6}\cmidrule(lr){7-10}
    Fault config. & Param.
      & P1 & P2 & P3 & P4
      & P1 & P2 & P3 & P4 \\
    \midrule
    \multicolumn{10}{@{}l}{\textit{Noise faults (Param.\ = SNR)}} \\
    $(V_z, V_a)$ & 35\,dB & Y & Y & Y & Y & Y & Y & Y & Y \\
    $(V_z, V_a)$ & 25\,dB & N & Y & Y & N & N & Y & Y & N \\
    $V_z$        & 25\,dB & N & Y & Y & N & N & Y & Y & N \\
    $V_a$        & 25\,dB & N & Y & Y & N & N & Y & Y & N \\
    $(V_z, V_a)$ & 15\,dB & N & N & Y & N & N & N & Y & N \\
    $V_z$        & 15\,dB & N & N & Y & N & N & N & Y & N \\
    \midrule
    \multicolumn{10}{@{}l}{\textit{Ramp-down faults (Param.\ = $\delta^{ramp}$)}} \\
    $V_z$ & $0.0025$ & N & Y & Y & N & N & Y & Y & N \\
    $V_z$ & $0.025$  & N & Y & N & N & N & Y & N & N \\
    $V_a$ & $0.23$   & N & Y & Y & N & N & Y & Y & N \\
    $V_a$ & $2.3$    & N & Y & N & N & N & Y & N & N \\
    $(V_z,V_a)$ & $(0.0025,0.23)$ & N & Y & Y & N & N & Y & Y & N \\
    $(V_z,V_a)$ & $(0.0025,2.3)$  & N & Y & N & N & N & Y & N & N \\
    $(V_z,V_a)$ & $(0.025,0.23)$  & N & Y & N & N & N & Y & N & N \\
    $(V_z,V_a)$ & $(0.025,2.3)$   & N & Y & N & N & N & Y & N & N \\
    \bottomrule
  \end{tabular}
\end{table}
\footnote{In our experiments $\delta_1 = \delta_4 = \delta$.
	Since $\tau_1 \le \tau_4$ and both properties use the same error band,
	P4 is implied by P1 in this instantiation, which explains the identical
	verdicts in Table~3. In general the two properties are independent
	(e.g., P1 does not imply P4 when $\delta_4 < \delta_1$).}
\paragraph{Analysis.}
SBSA and MBSA produce identical verdicts across both fault types and all
configurations, confirming the consistency of the two approaches.
Throughout, settling time (P1) and steady-state error (P4) are the most
sensitive properties and overshoot (P2) and rise time (P3) the most
robust, as the transient shape tolerates sensor degradation better than
long-term accuracy.  Under \emph{noise}, all properties hold at
$35$\,dB; P1 and P4 fail at $25$\,dB (independently of the affected
channel) and P2 at $15$\,dB, while P3 persists because the noisy
oscillations cross the $90\%$ threshold early.  Under \emph{ramp-down},
the persistent drift violates P1 and P4 at every rate and P2 always
holds (a monotone drift excites no overshoot peak); P3 is the only
rate-dependent property, holding at the slow rates and failing once
either channel runs at the fast rate, which delays the $90\%$ crossing
past the rise-time window.

\subsection{TMR Model Results}
\label{sec:results-tmr}

Table~\ref{tab:tmr} reports MBSA results for the TMR model under
fault multiplicities $P_{[0]}$ and $P_{[1]}$, with noise and
ramp-down faults applied simultaneously to the $V_z$ and $V_a$ sensor
groups.  The noise level is SNR~$= 25\,\mathrm{dB}$ ($k=10$); the
ramp-down rates are the slow rates $\delta_1^{ramp}$: $0.0025\,t$
for $V_z$ ($\tfrac{1}{1000}\times 2.5\,\mathrm{m/\,s}$) and
$0.23\,t$ for $V_a$ ($\tfrac{1}{1000}\times 230\,\mathrm{m/\,s}$).

\begin{table}[H]
	\centering
	\caption{TMR model property satisfaction under $P_{[0]}$ and
		$P_{[1]}$, MBSA (COMPASTA), at SNR~$=25\,\mathrm{dB}$ and slow
		ramp-down rates ($\delta_{V_z}^{ramp}=0.0025\,t$,
		$\delta_{V_a}^{ramp}=0.23\,t$).
		\textbf{Y} = satisfied; \textbf{N} = violated.}
	\label{tab:tmr}
	\small
	\setlength{\tabcolsep}{5pt}
	\begin{tabular}{@{}lccccccccc@{}}
		\toprule
		& & \multicolumn{4}{c}{\textbf{Noise}}
		& \multicolumn{4}{c}{\textbf{Ramp-down}} \\
		\cmidrule(lr){3-6}\cmidrule(lr){7-10}
		Fault class & Faulty replicas & P1 & P2 & P3 & P4 & P1 & P2 & P3 & P4 \\
		\midrule
		$P_{[0]}$ & 0 & Y & Y & Y & Y & Y & Y & Y & Y \\
		$P_{[1]}$ & 1 & Y & Y & Y & Y & Y & Y & Y & Y \\
		\bottomrule
	\end{tabular}
\end{table}

\paragraph{Analysis.}
Under both $P_{[0]}$ (no fault) and $P_{[1]}$ (one faulty replica per
group) the TMR model satisfies all four properties under noise and
ramp-down alike: with two healthy readings per group the majority voter
selects the correct value, making the fault invisible to the control
loop and reproducing the fault-free nominal behavior the masking
guarantee of Sect.~\ref{sec:modeling-voter}.  Beyond $P_{[1]}$ the voter
can no longer reconstruct the measurement and the fault propagates.
For ramp-down, this reduces the system to the nominal cases of
Sect.~\ref{sec:results-nominal}. For noise, the voter averages the
faulty replicas, which only reduces the noise variance: $P_{[2]}$ and
$P_{[3]}$ are statistically equivalent to the nominal noise case at a
higher SNR. We don't report them, since they characterize the controller's noise
tolerance rather than the voter.

\subsection{Discussion of the Results}
\label{sec:discussion}
The experimental results reveal several consistent trends across the
analyzed fault models. First, SBSA and MBSA produce identical verdicts
for all nominal-model experiments, confirming the consistency of the
two verification approaches despite their different modeling and
execution frameworks.

Regarding requirement sensitivity, settling time (P1) and steady-state
accuracy (P4) are consistently the most vulnerable properties, whereas
overshoot (P2) and rise time (P3) are generally more robust. This
suggests that sensor degradations affect long-term convergence more
severely than transient behaviour.

For additive noise faults, all properties are satisfied at the nominal
noise level (35 dB SNR). At 25 dB, P1 and P4 become violated, while P2
and P3 remain satisfied, independently of the affected sensor
channel. At more severe noise levels (15 dB and below), P2 is also
violated, leaving P3 as the only property that still holds. Noise
therefore causes a progressive degradation of controller performance,
first impacting convergence and steady-state behaviour and only later
affecting overshoot.
Ramp-down faults exhibit a different pattern. P1 and P4 are violated
for all analyzed degradation rates, while P2 remains satisfied in
every configuration. The rise-time requirement P3 is the only
rate-dependent property: it holds for slow degradations and is
violated when at least one channel experiences the faster ramp-down
rate.

The TMR experiments confirm the expected fault-masking capability of
the redundant architecture. For both noise and ramp-down faults, all
requirements remain satisfied when at most one replica per sensor
group is faulty, as majority voting prevents fault propagation to the
control loop. Beyond this threshold, the system is expected to exhibit
the same qualitative behaviors observed in the corresponding
nominal-model fault scenarios.
Overall, the results show that the considered sensor faults primarily
compromise convergence and steady-state performance, while
transient-response properties are comparatively resilient. They also
demonstrate the effectiveness of TMR in preserving all analyzed
requirements within its intended fault-tolerance assumptions.

The experimental results demonstrate that the MBSA approach achieves 
identical verdicts to SBSA while offering tangible advantages.

The \emph{coverage specification} differs between the two approaches.
SBSA relies on predefined metrics. In contrast, MBSA allows the user
to define customizable coverage criteria, which are then automatically
enforced by the model checker during scenario generation. This
flexibility enables systematic exploration of the fault space (and
state space) according to user-defined objectives.

The \emph{manual instrumentation effort} is also substantially
reduced.  SBSA requires manual encoding of temporal properties and
manual construction of the corresponding monitor, one for each
property, in the Simulink model, which may be time-consuming and
error-prone for complex properties.

Alternatively, using a custom toolbox~\cite{donze2010breach} it is
possible to specify properties in STL logic natively, mapping
individual sub-formulae to signals in the Simulink model and then run
a test campaign over a ranges of values for chosen parameters
(e.g. noise power factor $k \in [1, 3162]$). This method is based on
parameter-range sampling, and can be coupled with falsification
strategies~\cite{annapureddy2011staliro,khandait2024archcomp}, based
on heuristics and a stochastic optimization algorithm, to find
possible violations. On the other hand, MBSA supports the native
specification of LTL properties as property patterns, which are
directly handled by the COMPASTA toolchain, and it fully automates the
synthesis of monitors for LTL properties. While MBSA does not support
heuristic strategies to search for violations, the definition of the
test scenarios is based on the specification of formal conditions that
must be satisfied by any matching execution, which is more general
than parameter sampling. Moreover, the coverage of the generated
scenarios with respect to their specification is formally guaranteed.

Finally, the \emph{fault definition} itself offers greater
flexibility in MBSA. In SBSA, faults are defined at the signal level,
attached to specific input or output signals of a block (e.g.,
Simulink fault analyzes applied faults to signals only). In MBSA,
faults are defined at the component level, allowing faults to be
injected directly into any system component, and may depend on the
interface of the component itself. This component-centric approach
simplifies fault specification and enables more complex fault models,
such as component degradation or custom internal failures.

\begin{table}[H]
	\centering
	\caption{Comparison of SBSA and MBSA verification efforts for the ROSACE benchmark.}
	\label{tab:effort-comparison}
	\small
	\setlength{\tabcolsep}{5pt}
	\begin{tabular}{@{}lcc@{}}
		\toprule
		\textbf{Metric} & \textbf{SBSA} & \textbf{MBSA} \\
		\midrule
		Temporal properties encoding/verification effort& high & low \\
		Coverage specification & Predefined metrics & Customizable \\
		Coverage of TMR fault configurations & Manual & Automated \\
		Automatic monitor synthesis for LTL& No & Yes\\
		\bottomrule
	\end{tabular}
\end{table}

\section{Conclusions and Future Work}
\label{sec:conclusions}
In this work we have extended the ROSACE case study
of~\cite{pagetti2014rosace} with a representative set of run-time
faults, and we have systematically evaluated the impact on the safety
and performance requirements of the system. Our approach combines
formal methods to automatically generate sets of test scenarios that
meet user-defined coverage criteria, with co-simulation techniques to
generate execution traces. We have formalized the requirements of
ROSACE in temporal logic and used formal techniques to validate them
against several variants of the original model (with and without
faults, with and without TMR redundancy). We have shown the advantages
of using formal methods and Model-Based Safety Analysis, compared to
traditional purely simulation-based approaches.  Our approach is
implemented on top of the TASTE tool by ESA, a comprehensive toolchain
targeted to the model-based development and implementation of
embedded, real-time systems. Although not addressed in this paper, as
part of our future work, we would like to investigate using TASTE for
the coding, deployment and validation of ROSACE on a real target, and
compare the results with~\cite{pagetti2014rosace}. Additional lines of
research include extending the ROSACE case study further, by
considering other types of faults or redundancies, such as, e.g.,
in~\cite{R-ROSACE,DDP15}. Finally, we will consider applying our
approach to other case studies, e.g.,
\cite{BW14,FMcasestudies,Picchiami24,2024sais.conf..297B}.

\bibliographystyle{eptcs}   
\bibliography{references}



	
\end{document}